# High Pressure and Road to Room Temperature Superconductivity

Lev P. Gor'kov[1] and Vladimir Z. Kresin[2]

[1]NHMFL, Florida State University, 1800 E. Paul Dirac Drive, Tallahassee, Florida, 32310, USA

[2]Lawrence Berkeley Laboratory, University of California,

1 Cyclotron Road, Berkeley, CA 94720, USA

Abstract

High pressure serves as a path finding tool towards novel structures, including those with very high Tc. The superconductivity in sulfur hydrides with record value of Tc =203 K (!) is caused by the phonon mechanism. However, the picture differs from the conventional one in important ways. The phonon spectrum in sulfur hydride is both broad and has a complex structure. High value of $T_c$ is mainly due to strong coupling to the high-frequency optical modes, although the acoustic phonons also make a noticeable contribution. New approach is described; it generalizes the standard treatment of the phonon mechanism and makes it possible to obtain an analytical expression for Tc . It turns out that, unlike in the conventional case, the value of the isotope coefficient varies with the pressure and reflects the impact of the optical modes. The phase diagram, that is the pressure dependence of Tc, is rather peculiar. A crucial feature is that increasing pressure results in a series of structural transitions, including the one, which yields the superconducting phase with the record Tc . In a narrow region near P≈150 GPa the critical temperature rises sharply from Tc ≈120K to Tc ≈200K. The sharp structural transition, which produces the high Tc phase, is a first-order phase transition caused by interaction between the order parameter and lattice deformations. A remarkable feature of the electronic spectrum in the high Tc phase is the appearance of small pockets at the Fermi level. Their presence leads to a two-gap spectrum, which can, in principle, be observed, with the future use of tunneling spectroscopy. This feature leads to non-monotonic and strongly asymmetric pressure dependence of Tc. Other hydrides can be expected to display even higher values of Tc, up to room temperature. The fundamental challenge lays in creation a structure capable of displaying high Tc at ambient pressure.

**CONTENTS**

## I. Introduction.

In a recent dramatic development, superconductivity with a critical temperature $T_c$=203 K (!) has been observed in sulfur hydride under high pressure (Drozdov, Eremets et al., 2015). This development is the most significant breakthrough since the discovery of the high $T_c$ oxides (Bednorz and Mueller, 1986). There is every reason to anticipate even higher values of $T_c$ for other hydrides, which means that achieving superconductivity at room temperature now appears perfectly realistic.

 We focus below on theoretical aspects, which can provide a clue to understanding the specifics of the new superconducting state. It should be emphasized that in many respects the system is quite unusual. As will be discussed below, the observed phenomenon can be explained by the phonon mechanism. Nevertheless, a number of key features reveal that the picture differs from the conventional one. In this Colloquium we devote particular attention to these features. One of them is the peculiar pressure dependence of the critical temperature. Not only does external pressure result in metallization and the appearance of superconductivity, but also the value of $T_c$ grows dramatically with a further rise in pressure. Moreover, this variation turns out to be non-monotonic. It also should be noted that while Cooper pairs formation is mediated by phonons, the complex structure of the phonon spectrum and its broad range (all the way up to $\Omega \approx 2000$ K, where $\Omega$ is the phonon frequency) makes it necessary to modify the conventional treatment. While the strong isotope effect affirms the action of the phonon mechanism, the value of the isotope coefficient turns out to vary with pressure. Below we discuss all these interesting aspects of the new development in superconductivity.

  Despite its rich and interesting prehistory, the discovery came as a surprise, especially for those who accept the notion that electron-phonon interaction cannot give rise to such a high $T_c$. The background of this notion will be discussed below (Sec. IIIA). One also should not lose track of the fact that this impressive discovery came about thanks to the remarkable progress in high-pressure technique and to experimental innovations by the M.Eremets group (Max Planck Institute, Mainz, Germany ). One should stress also that the study of the hydrides has attracted an attention of many theoretical groups (see below,Ch.3). One should give a special credit to the group of Y.Ma and also to D.Duan, T.Cui and their collaborators (Julin University, Changchun, China) (see below,Chs.III,VII). Their remarkable studies brought a special attention to the sulfur hydrides and motivated the key experimental studies.

A few remarks on the historical perspective are in order. The phenomenon of superconductivity was discovered more than one hundred years ago by H. Kamerlingh Onnes (1911). While measuring the temperature dependence of the electrical resistance $R(T)$ of mercury he observed that at the temperature of 4.2 K the resistivity suddenly vanishes. The dissipationless ($R=0$) state, which emerged, was named the superconducting state. Subsequently superconductivity was discovered in many other materials. Moreover, it was soon realized that the loss of resistance was only one facet of the superconducting state, hence the latter corresponded to a qualitatively new state of matter. Its most fundamental feature, the so-called Meissner effect ( Meissner and Ochsenfeld, 1933), is manifested in the expulsion of the magnetic field from the bulk of the sample (anomalous diamagnetism).

The microscopic theory of superconductivity was created by Bardeen, Cooper and Schrieffer (BCS) in 1957, almost 50 years after the experimental discovery. According to the BCS theory, the key microscopic factor behind the phenomenon is the attraction between electrons mediated by the exchange of phonons, such that below $T_c$ within the electronic system there forms a macroscopic manifold of bound electron pairs (known as the Cooper pairs). Thus the attraction has its origin in the ionic system; qualitatively the effect may be related to the dielectric function changing its sign at low frequencies.

The well-known expression for the critical temperature is

$$T_c \approx \tilde{\Omega} \exp\left(-\frac{1}{\lambda - \mu^*}\right) \qquad (1.1)$$

Here $\tilde{\Omega}$ is the characteristic phonon frequency (typically on the order of the Debye temperature), $\lambda$ is the electron-phonon coupling constant, and $\mu^*$ is the so-called Coulomb pseudopotential which characterizes the direct electron-electron repulsion (usually $\mu^* \approx 0.1$-$0.15$). Eq.(1.1) is valid in the weak coupling approximation ($\lambda \ll 1$). Note that the specific value of the pre-exponential factor is determined by the renormalization effect (see below, Sec.IIIA).

The search for superconducting materials with higher critical temperatures has been ongoing. Fig.1 shows how the maximum transition temperatures within different superconductor families has grown with time. In the first 75 years progress was rather modest (from $T_c = 4.2$ $K$ for mercury up to $T_c \approx 23$ K in $Nb_3Ge$). A breakthrough came in 1986 when Bednorz and Müller discovered a new family of superconducting materials, the copper oxides (cuprates) and observed a $T_c$ close to 40 K in the La-Sr-Cu-O compound. Subsequent research on cuprates raised their $T_c$ all the way up to 130 $K$ in the HgBaCaCuO compound (Schilling et al., 1993). For what follows, it is noteworthy that under pressure $T_c$ was raised up to $\approx 160$ $K$ (Gao et al., 1994). Until very recently, this remained the highest critical temperature ever observed. (Fig.1)

II. **Hydrides**

  A. **Metallic hydrogen.**

As noted in the Introduction, recent work on sulfur hydrides under high pressure led to the observation of a superconducting state with the record value of $T_c=203K$. This achievement was preceded by developments, which started almost fifty years ago (Ashcroft, 1968). According to the BCS theory, $T_c$ is proportional to the characteristic phonon frequency $\tilde{\Omega} \propto 1/\sqrt{M}$ [see Eq.(1.1)]. One may expect, therefore, that metallic hydrogen should have a high value of $T_c$ : for such a light metal the characteristic phonon frequency in the prefactor of Eq.(1.1) is high and (ignoring for the moment the magnitude of the exponential factor) $T_c$ may also turn out to be rather high. However, this prediction can be verified only under very high pressure. Indeed, hydrogen first must become metallic, but the transition from the molecular phase into the metallic state is known to require high pressure (Wigner and Huntington,1935). This is the dissociative transition to an atomic lattice, which occurs through compression of solid molecular hydrogen.

A conductive (probably, semimetalic ) state of hydrogen was observed by Eremets and Troyan (2011) at room temperature under the pressure of 260-270 GPa (recall that 100 GPa corresponds to a million atmospheres).

The phase diagram for hydrogen is rather complicated (Fig. 2), and the determination of this diagram was a non-trivial task. The main complication derives from the fact that the usual technique for the structure determination, which underlies the phase diagrams of materials, namely X-ray diffraction, is not conclusive because scattering by hydrogen is very weak.

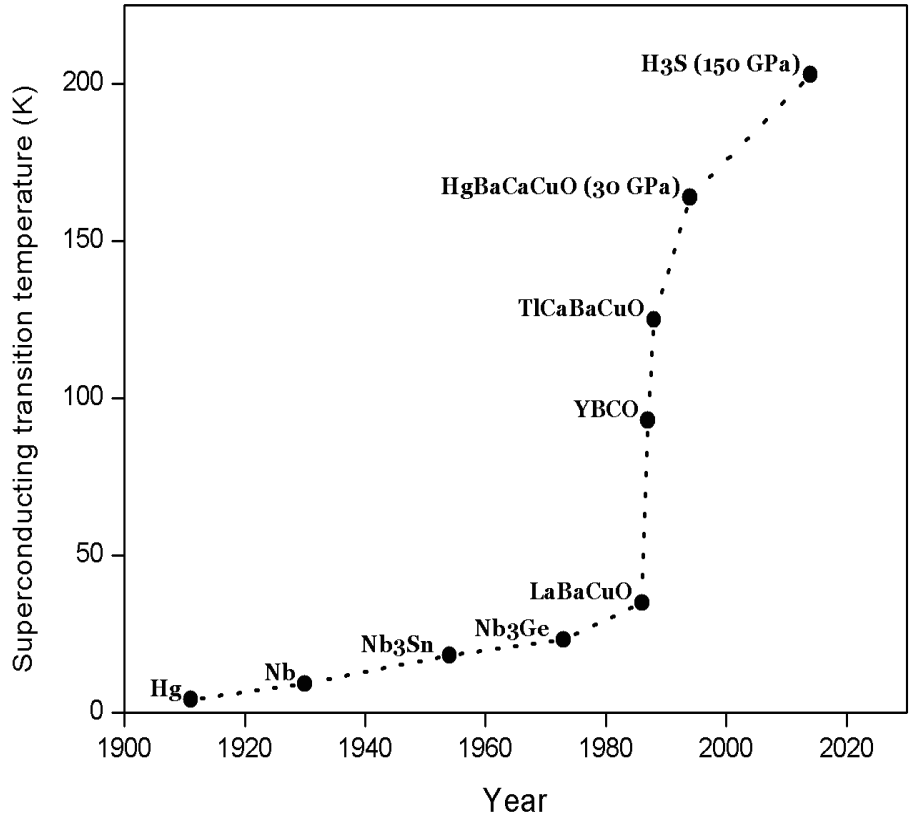

Fig.1. Increase in $T_{c;max}$ with time. During the next 75 years since the discovery the increase was $\Delta T_c \approx 18K$ ; the main focus was on the Nb-based materials. The highest $T_c$ for the cuprates at ambient pressure ($T_c \approx 130K$) was raised under the pressure up to $T_c \approx 160K$. At present, the curve is extended up to $T_{c;max} = 203K$ observed for sulfur hydride under high pressure.

The diagram in Fig. 2 is based on Raman and infrared measurements along with resistivity data. The low temperature phases I-III wer observed at relatively low pressures (≤150 GPa, see the review by Mao et al.,1994).

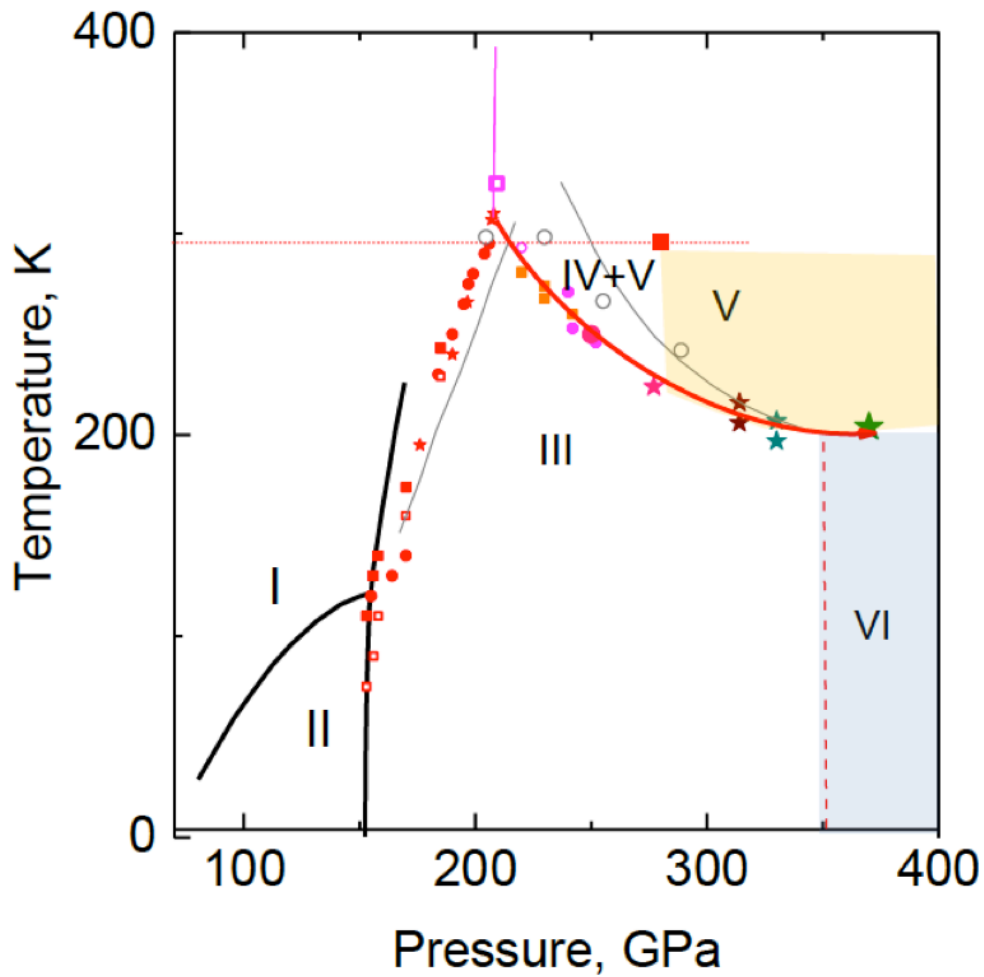

Fig. 2. The phase diagram of hydrogen. Letters I-VI indicate domains for different phases. Phases I-III correspond to molecular state of pure hydrogen. Their boundaries can be determined from Raman and infrared measurements. Phase IV is mixed atomic-molecular phase, phase V-atomic liquid. Recently observed phase VI corresponds to the metallic state. From Eremets et al., 2016.

Recently Eremets et al. (2016) observed a new phase (phase VI in Fig.2) at pressures ≳ 360 GPa and temperatures <200 *K*. This phase displays a drastic drop in resistivity, characteristic featureless Raman spectra, and the absence of a photoconductive response. These properties are characteristic of the metallic state.

As concerns the superconducting state of metallic hydrogen, theoretical calculations (McManon et al.,2012) show that a high-$T_c$ state of pure metallic hydrogen requires pressures on the order of ≈500 GPa. Hopefully, this phenomenon will be observed in the near future.

### B. **Hydrides. High $T_c$ superconductivity.**

The reasoning for an elevated value of $T_c$ outlined above applies not only to pure hydrogen but also to hydrides, that is, to materials containing hydrogen as one of their components (Ashcroft, 2004; Wang and Ma, 2014). The presence of hydrogen results in the appearance of high-frequency optical phonon modes, whereas heavy ions provide additional acoustic modes, which also contribute to the pairing. At the same time, the metallization of such compounds does not require extraordinarily high pressures. In fact, hydrides are even more promising materials than pure metallic hydrogen. Indeed, the presence of more than one ion in the unit cell, in this case an additional hydrogen ion, leads to an appearance of high frequency optical modes which, in addition to high frequency , are characterized by a high density of states. The latter is beneficial for superconductivity.

A number of density-functional theory studies supported the high promise of hydride compounds. For example, calculations (Gao et al., 2008) suggested that $GaH_3$ at P≈160 GPa would display $T_c$≈140 K and $Si_2H_6$ at ≈275 GPa would display $T_c$≈73 K – 86 K  (Jin et al.,2010) .The most thermodynamically stable structures were established by calculating the enthalpy- difference curves.  Among others, the minima hopping method was employed  (Goedeker, 2004).

   Initial experiments on $SiH_4$ ( Eremets et al.,2008) demonstrated that hydrides can indeed support a superconducting state, although the critical temperature was a relatively modest ≈17K.  Later, following the discovery of the record high $T_c$  in sulfur hydrides, a high $T_c$  state    (≈100 K) in covalent hydride phosphine (P-H) was also observed (Drozdov et al., 2015). Theoretical analysis (Flores-Divas et al, 2016) suggests that the $PH_{1,2,3}$ systems indeed have a rather high $T_c$ , but the material is probably, in a metastable state.

   In 2014 Li et al. made the remarkable prediction that metallic sulfur hydride would become superconducting with $T_c$ ≈80 K under the relatively low pressure *P≈100 GPa* .  Following the first experimental observation of such a superconducting state, Eremets and his collaborators continued increasing the pressure and discovered that $T_c$  goes up significantly all the way to $T_c$≈*203 K* (Drozdov, Eremets et al., 2014, 2015a; see also the review by Eremets and Drozdov, 2016). Such a remarkable observation was explained by mixed valence of S and formation of sulfur hydride with higher hydrogen content. This assumption was in agreement with very interesting independent theoretical study by Duan et al (2014) More specifically, increase in pressure is accompanied by the formation of H$_3$S structural units (Fig.3a) via the transformation:

$$3 H_2S \rightarrow 2H_3S + S \qquad (2.1)$$

The transformation $H_2S \rightarrow H_3S$ also has been confirmed by detailed calculations by Bernstein et al.,2015 and Errea et al.,2015.

Sulfur hydride at the pressure 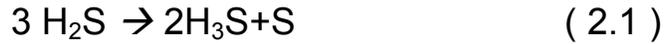 $P \approx 90 GPa$ has $T_c \approx 90 K$. As mentioned above, the subsequent increase in pressure leads to a large increase in the value of the critical temperature up to $T_c \approx 200$ K. This pressure dependence implies that the presence of hydrogen is necessary but not sufficient by itself for reaching the highest $T_c$ values. Indeed, the sample at 90 *GPa* contains hydrogen and consequently high frequency modes. The $T_c$ of 90 *K* is very high and would be sensational 30 years ago, but it is much below the 203 *K* that is achieved under higher pressure. Therefore the pressure increase brings in some additional factors. As mentioned above, Drozdov, Eremets et al. (2014,2015a) suggested that the rise in $T_c$ is due to the formation of new compounds with higher valence states of sulfur.

As will be described below, the pressure increase changes the crystal structure of sulfur hydride, thereby signifying a structural phase transition. The structural transformation plays the the crucial role in the observed behavior.(Fig.3).

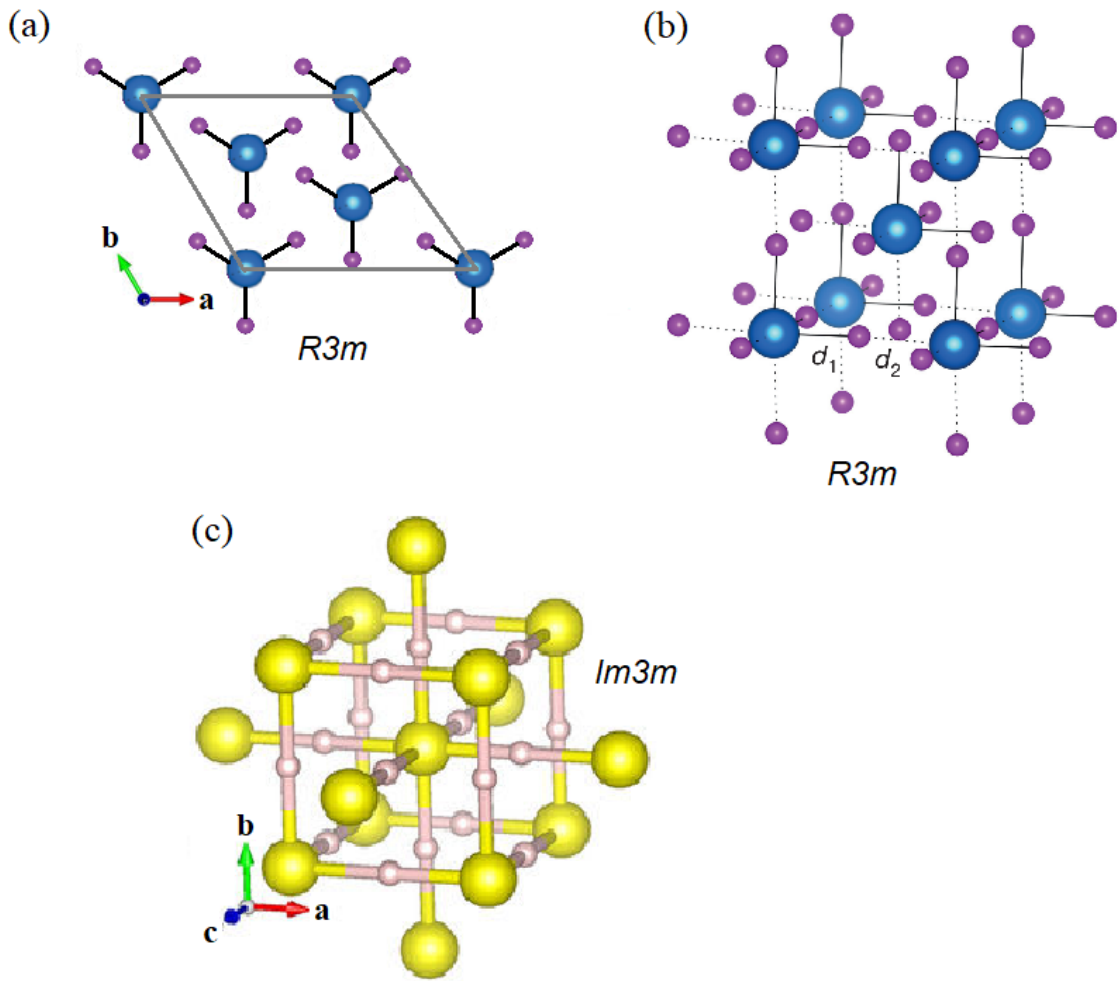

Fig. 3. Structures of the "low $T_c$" (R3m; $T_c \approx 120K$) and "high $T_c$" ( Im-3m; $T_c \approx 200K$) phases: (a) structure of the R3m phase (top view); one can see the $H_3S$ entities. Large (small) spheres denote sulfur (hydrogen) ions. After Duan et al.,2014 (b) structure of the R3m phase (side view); one can see that $d_1 \neq d_2$, $d_i$ (i=1.2) are the distances between the H ion and neighboring S-ions .From Errea et al.,2016.  (c) structure of the Im-3m phase . Unlike the "low $T_c$" phase, the distances $d_1$ and $d_2$ are equal in the cubic high $T_c$ phase Im-3m. From Duan et al.,2014 .

Below we focus on the two phases with the highest values of $T_c$ (Fig.3) . One of them (Figs.3a,b) has the crystal structure ,which corresponds to *R3m* symmetry (see, e.g., Massa, 2004). Increasing the pressure further leads, at P≈150 GPa, to the appearance of a different phase (Fig.3c). It is this new structure that displays the record- high $T_c$ .The symmetry group of the lattice is cubic $O_h$ (*Im-3m*). The theoretically predicted structure is in agreement with X-rays data (Einaga et al.,2016; Goncharov et al.,2016;2017). The usual X-rays spectroscopy does not allow to determine the structure of the hydrides with a high accuracy, because the light H ions do not provide strong scattering. In addition, the multiphase nature of the studied samples (Goncharov et al., 2016) also represents a serious complication. Recently, advance spectroscopy has been employed to monitor how the structure evolves with pressure (Goncharov et al., 2017). More specifically, the synthesis performed out of S and molecular hydrogen along with cyclotron XRD technique and Raman spectroscopy did allow the researchers to study all structures formed with an increase in pressure. It turns out that the transformation (2.1) occurs at pressure *P*> 40 GPa . It has been also demonstrated that the phase at *P*> 110GPa ($T_c$≈120 K) has R3m symmetry (Figs.3 a,b). As for the most interesting high $T_c$ phase, its structure (Fig. 3c) , indeed, is characterized by the Im-3m symmetry.

The transformation (2.1) supports the suggestion by Drozdov et al. (2015, see above) about the mixed valence state of sulfur. The latter factor turns out to be essential (see Sec. IVC below).

**C. Superconductivity in hydrides: Main properties.**

Experimentally, the onset of the high $T_c$ superconducting state in sulfur hydrides ( Drozdov et al.,2015) is detected by the drastic resistance drop near $T_c$ . A sharp transition was observed in annealed samples. The measured resistance was at least two orders of magnitude below that of pure copper.

The critical temperature shifts downward in the presence of an external magnetic field. Magnetic susceptibility measurements reveal an abrupt transition into the diamagnetic state (the Meissner effect). This key result was also confirmed by a direct observation (Troyan et al.,2016) of magnetic field expulsion as detected by the response of a thin Sn film placed inside the bulk sample. The Meissner effect was also observed later by Huang et al. (2016) by means of AC magnetic susceptibility measurements.

Below we concentrate on the two phases with the highest $T_c$. One of them (R3m; Fig.3a,b) has $T_c \approx 120\ K$ , and the other (*Im-3m*; Fig.3c) has the highest $T_c$ of ≈200 K. Below, we will refer to the former structure as the "low $T_c$ phase" (although this name sounds ironic for $T_c \approx 120\ K$), and to the second structure as the "high $T_c$ phase."

The question of the mechanism of superconductivity with such a record $T_c$ is of fundamental interest. The strong isotope effect indicates that pairing is provided by phonon exchange. The main contribution comes from the high-frequency optical modes. Nevertheless, as stressed in the Introduction, the picture is far from conventional.

The phase diagram, that is, the pressure dependence of $T_c$ is very peculiar. Indeed, Fig. 4 demonstrates that $T_c$ is strongly dependent on the applied pressure (Einaga et al, 2016). It increases from ≈100 – 120 *K* up to the record ≈200 *K* over the relatively narrow pressure interval 125-150 *GPa*. We argue below that such a rapid $T_c$ change is a fingerprint of a first-order structural transition.

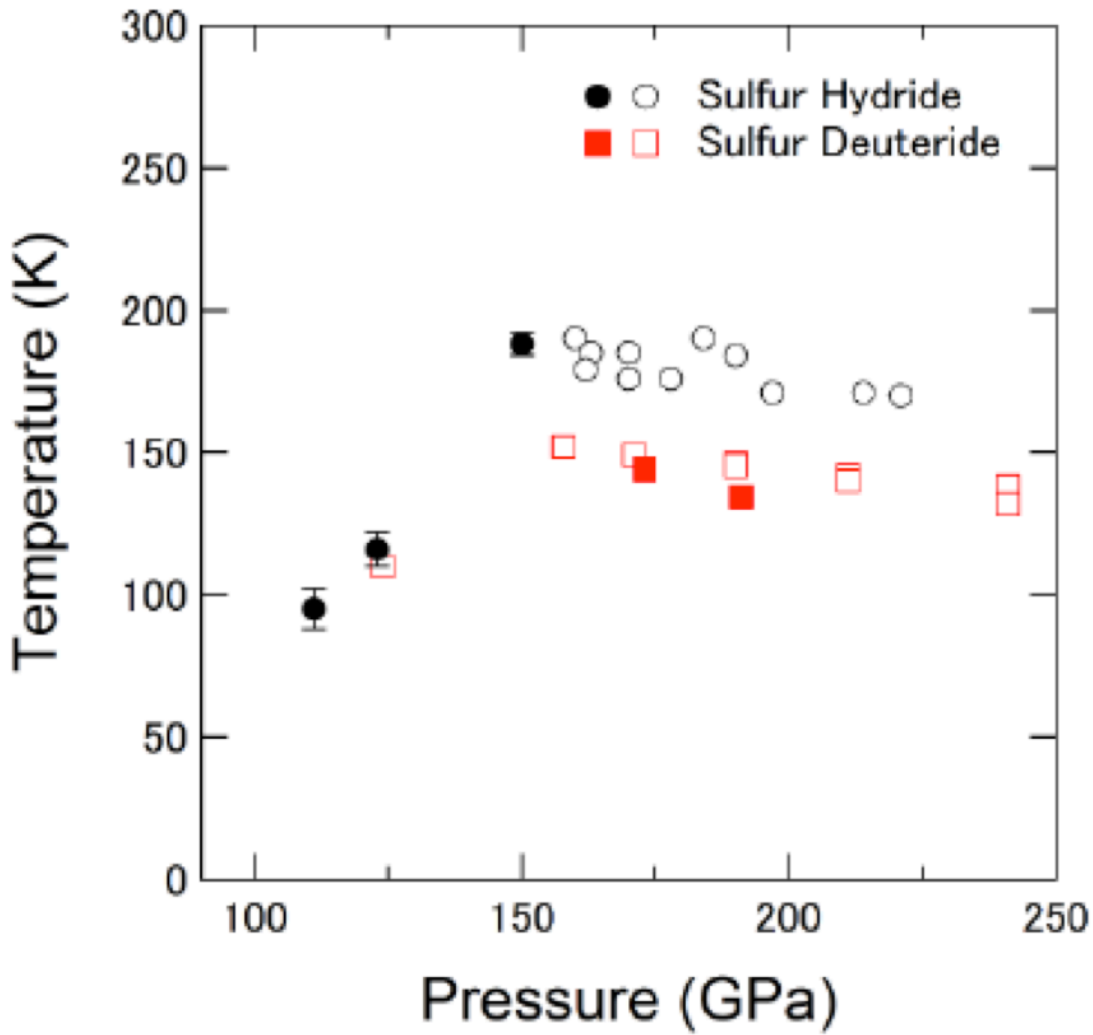

Fig. 4. Pressure dependence of $T_c$. The data for annealed samples are presented. One can see a large increase in the value of $T_c$ in the region near *P*=140GPa. Adapted from Einaga et al.,2016.

Note that once $T_c$ reaches its maximum value, it decreases upon further rise in applied pressure. The decline is rather slow, so that the $T_c$ dependence is strongly asymmetric relative to $T_{c;max}$. Such a dependence is unusual and requires an explanation. We interpret this feature (see Sec. VIID) as deriving from the impact of the superconducting order parameters on the main Fermi surfaces and on small pockets, with the latter appearing in the electronic spectrum of the high $T_c$ phase in the course of a first-order phase transition.

Note also that because of the complex structure and the width of the phonon spectrum evaluation of $T_c$ and of the isotope coefficient (whose value turns out to be pressure-dependent) must be carried out with considerable care.

The next chapter contains a general description of the phonon mechanism with emphasis on the strong coupling case. Subsequent chapters focus on the challenges described above.

### III. Electron-phonon interaction. Critical temperature.
#### A. Main equations. Coupling constant and $T_c$.

Our main goal is to evaluate the value of the critical temperature for the hydrides. The usual BCS model is not applicable here, because it has been developed in a weak coupling approximation ( $\lambda$<<1 and correspondingly, $2\pi T_c << \tilde{\Omega}$ ). The equations describing strong-coupled superconductors (Migdal,1958; Eliashberg,1960) contain the so-called phonon propagator (see below, Eq.(3.1), and therefore, the phonon frequency $\Omega$. The equation for the pairing order parameter $\Delta(\omega_n)$ has the form:

$$\Delta(\omega_n)Z = \pi T_c \sum_m \int d\Omega \frac{\alpha^2(\Omega)F(\Omega)}{\Omega} D(\tilde{\Omega}, \omega_n - \omega_m) \frac{\Delta(\omega_m)}{|\omega_m|} \quad (3.1)$$

Here

$$D = \frac{\Omega^2}{\Omega^2 + (\omega_n - \omega_m)^2} \quad (3.2)$$

is the phonon propagator, $\Omega$ is the phonon frequency, $\omega_n=(2n+1)\pi T_c$ (the method of thermodynamic Green's functions is employed; see, e.g., Abrikosov et al.,1975 ); one should add also the Coulomb pseudopotential $\mu^*$. The factor $Z$ is the renormalization function determined by the relation:

$$Z = 1 + (\pi T_c / \omega_n) \sum_m \int d\Omega \frac{\alpha^2(\Omega)F(\Omega)}{\Omega} D(\tilde{\Omega}, \omega_n - \omega_m) \frac{\omega_m}{|\omega_m|} \quad (3.3)$$

The renormalization function describes the "dressing" of electrons moving through the ionic lattice.

Eqs.(3.1),(3.3) contain the important quantity, the function $\alpha^2(\Omega)F(\Omega)$. Here $F(\Omega)$ is the phonon density of states, $\alpha^2(\Omega)$ describes the electron-phonon interaction and contains the corresponding matrix element (See e.g., Scalapino,1969; Grimvall,1981; Wolf,2012).

In addition, one can introduce an important parameter, so-called coupling constant $\lambda$, defined by the relation:

$$\lambda = \int d\Omega \frac{\alpha^2(\Omega)F(\Omega)}{\Omega} \quad (3.4)$$

Note that Eqs. (3.1), (3.3) do not explicitly contain the coupling constant λ. Indeed, generally speaking, this constant cannot be factored out, because the phonon frequency enters not only in the factor $\alpha^2(\Omega)F(\Omega)$ but also in the phonon propagator D($\omega_n$-$\omega_m$), which, in addition, depends on $\omega_n$-$\omega_m$ (see Eq.(3.2) ). It is apparent from Eqs. (3.1), or (3.3 ) that the coupling constant can be factored out, if Eq. (3.1 ) does not contain a phonon propagator function (e.g., D≈1 for the weak coupling case) or if the phonon propagator $D$ slowly depends on the frequency Ω, so that Ω in D can be replaced by its average value. When the function $\alpha^2(\Omega)F(\Omega)$ is known, the value of the critical temperature can be evaluated from the non-linear equation which looks like (3.1) with the replacement:

$|\omega_n| \mapsto [\omega_n^2 + \Delta^2(\omega_n)]^{1/2}$ in the denominator of the integrand. Such an equation describes the order parameter at any temperature ( then $\omega_n$=(2n+1)πT).The calculation can be performed numerically ,without invoking the coupling constant concept. We will discuss the corresponding method below (Sec. IVA) while focusing on the sulfur hydrides.

At the same time the possibility to introduce the coupling constant, λ ,is very beneficial for the analysis. The concept of coupling constant is commonly used to study usual superconductors. It allows one to deduce the analytical expressions for $T_c$ and interpret its dependences on other parameters in the problem. Such an approach is justified, because usually the function $\alpha^2(\Omega)F(\Omega)$ is characterized by the peak structure in phonon density of states F($\Omega$) (see, e.g.[1] Wolf ,2012, and also Fig.5b ). The latter structure corresponds to the short-wavelength part of the spectrum where the mode dispersion is weak. The phonon propagator changes slowly on the scale corresponding to the peak structure, and this permits the replacement of $\Omega$ in the phonon propagator by its average value $\tilde{\Omega}$ ; the latter can be taken either as $\tilde{\Omega}=<\Omega^2>^{1/2}$ (see, e.g., Louie and Cohen ,1977, and the reviews by Grimvall ,1981;  Kresin and Wolf (2009)) , or as $\tilde{\Omega}$ =< $\Omega_{log}$>which is close to $\tilde{\Omega}=<\Omega^2>^{1/2}$  (Allen and Dynes ,1975; Carbotte,1990) . The average $<f>$ is defined by the relation:

$$<f(\Omega)>=(2/\lambda)\int d\Omega f(\Omega)\alpha^2(\Omega)F(\Omega)\Omega^{-1}$$ . If $\tilde{\Omega}=<\Omega_{log}>$,then f= log

. Below we have chosen $\tilde{\Omega}=<\Omega^2>^{1/2}$, so that

$$<\Omega^2>=(2/\lambda)\int d\Omega \Omega \alpha^2(\Omega)F(\Omega)$$; $\lambda$ is defined by Eq.(3.4). As a result, Eqs.( 3.1 ),(3.3) can be written in the form

$$\Delta(\omega_n)Z = \pi T_c \lambda \sum_m D(\tilde{\Omega},\omega_n - \omega_m)\frac{\Delta(\omega_m)}{|\omega_m|} \qquad (3.5)$$

$$Z = 1+(\pi T_c/\omega_n)\lambda \sum_m D(\tilde{\Omega},\omega_n - \omega_m)\frac{\omega_m}{|\omega_m|} \qquad (3.6)$$

D is defined by Eq.(3.2) and

$$\tilde{\Omega}=<\Omega^2>^{1/2} \qquad (3.7)$$

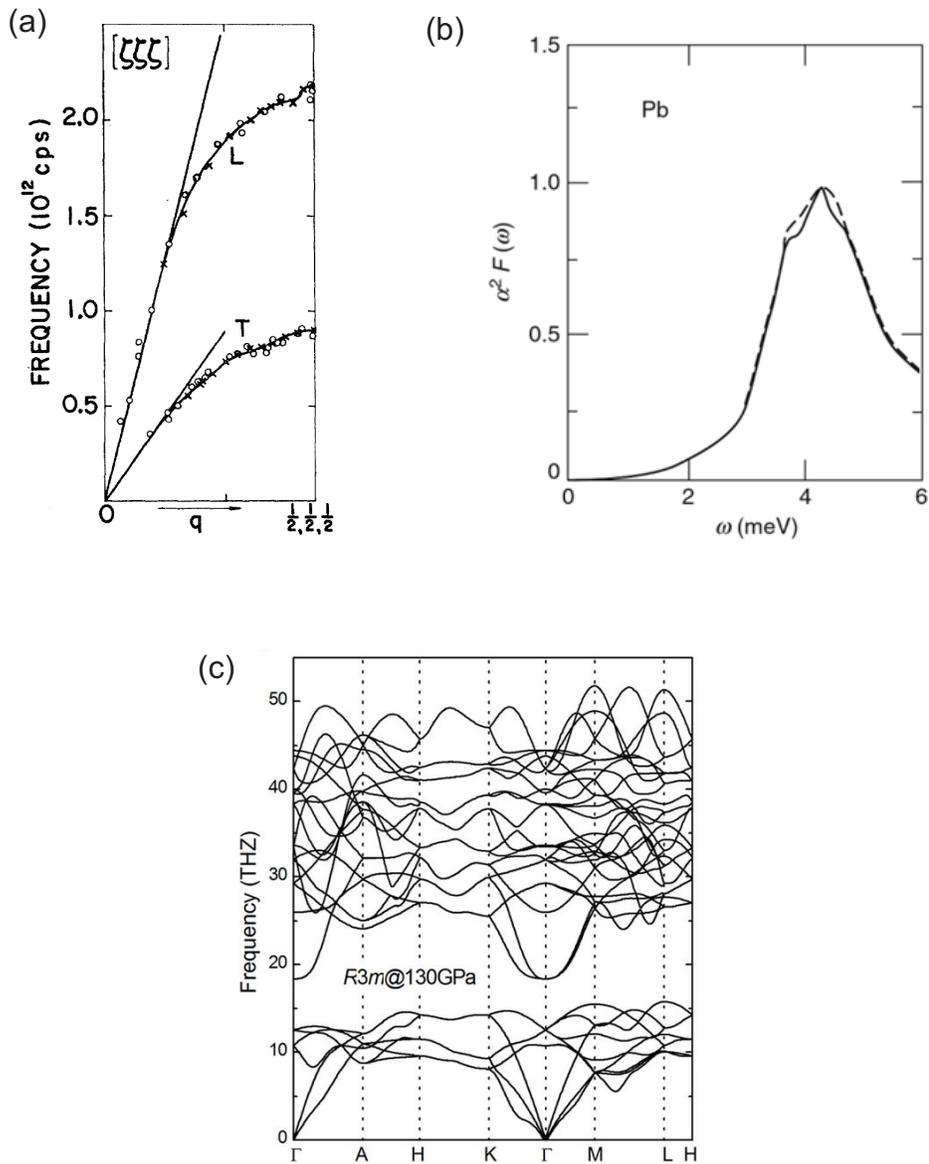

Fig.5.a) Phonon spectrum of Pb ; $\Omega$ and q are the phonon frequency and momentum, from Brockhouse et al. ,1962; b) the function $\alpha^2(\Omega)F(\Omega)$ for Pb. Adapted from McMillan and Rowell,1965; c) phonon spectrum of sulfur hydride (R3m phase; from Duan et al.,2014)

As is known, the solution of the equation (3.5) can be written in analytical form and the explicit expression depends on the strength of the electron-phonon interaction (see the review by Kresin et al.,2014). For the weak coupling case ($\lambda \ll 1$; then $2\pi T_c \ll \tilde{\Omega}$), $D \approx 1$, $Z \approx 1$, and we obtain a well-known BCS expression, Eq.(1.1). For the intermediate coupling one should take into account the renormalization of the coupling constant, since $Z \approx 1+\lambda$, and $T_c$ is described by the corresponding equation (Nakajima and Watabe ,1963, see Grimvall, 1977). Note that even in the weak coupling case the renormalization function affects the value of the pre-exponential factor (Karakosov et al,1976; Wang and Chubukov,2013), which appears to be equal to $a \approx 0.25\tilde{\Omega}$, see below Eqs.(4.5),(7.1). A well-known McMillan-Dynes expression (McMillan ,1968; Dynes ,1972) is widely used in literature:

$$T_c = \left(\tilde{\Omega}/1.2\right)\exp\left[-\frac{1.04(1+\lambda)}{\lambda - \mu^*(1+0.62\lambda)}\right] \quad (3.8)$$

This expression is valid for $\lambda \lesssim 1.5$. Note that for small $\lambda$ Eq. (3.8) can be written in the form similar to Eq. (1.1) with $a \approx 0.3\tilde{\Omega}$, which is close to the value mentioned above. For larger values of $\lambda$ one should use the different expression, which was obtained by numerical modification of Eq. (3.8) (Allen and Dynes, 1975), see below, Sec.IVA. One can use also the expression valid for any $\lambda$ (Kresin, 1987), see below, Eq. (7.1). Note that for the case of very strong coupling ($\lambda \gtrsim 5$; then $\pi T_c \gtrsim \tilde{\Omega}$) the dependence of $T_c$ on $\lambda$ is entirely different from Eqs. (1.1), (3.8) (we omitted μ* for simplicity) and has a form:

$$T_c = 0.18\lambda^{1/2}\tilde{\Omega}$$

(3.9)

(see Allen and Dynes, 1975; Kresin et al.,1984) .

In addition, one can introduce the following important relation (McMillan,1968):

$$\lambda = <I>^2 \nu / M\tilde{\Omega}^2 \qquad (3.10)$$

(I describes the electron-phonon scattering, $\nu$ is the electronic density of states)

The important question is whether there exists an upper limit of the value of $T_c$ . A "myth", which persists even today, claims that the phonon mechanism is unable to provide the values of the critical temperature higher that ~30K. For some scientists the argumentation is based on Eqs. ( 3.8),(3.10 ). Indeed, neglecting $\mu^*$ for simplicity and calculating $\partial T_c / \partial \tilde{\Omega}$, one can easily find the maximum value of $T_c$ ; this value corresponds to $\lambda$ 2. However the McMillan equation is valid only for $\lambda \gtrsim 1.5$ . This limitation was mentioned above, below Eq.(3.8). In connection with this, let us note that the MacMillan-Dynes equation (3.8) was obtained by taking into consideration the renormalization function Z≈1+λ ( see Eqs.(3.1) and (3.3)) and by fitting the coefficients to describe the data on Nb. The same limitation was obtained by Geilikman et al (1975), who derived analytically the equation similar to (3.8) ,see below, Eq.(4.6);the derivation is valid for $\lambda \gtrsim 1.5$ (then $(\pi T_c/\tilde{\Omega})^2$<<1. Therefore, the value $\lambda$  2 is outside the range of the  applicability of the MacMillan-Dynes equation. One can see from Eq. ( 3.9) that the mentioned upper limit for $T_c$ does not exist.

Another erroneous restriction was imposed not on the dependence of  $T_c$ on $\lambda$ (see above), but on the limiting value of the coupling constant $\lambda$ itself. In the framework of the so-called Froelich Hamiltonian : $\hat{H} = \hat{H}_{el} + \hat{H}_{ph} + \hat{H}_{int}$ ,where $\hat{H}_{ph}$ contains the phonon frequency $\Omega_0(q)$ and $\hat{H}_{int}$ describes the electron-phonon interaction, one obtains for the frequency $\tilde{\Omega}$ ,renormalized by the electron-phonon interaction, the expression: $\tilde{\Omega} = \Omega_0(1-2\lambda)^{1/2}$ .One would conclude that the lattice becomes unstable at value of the coupling constant    =0.5, and therefore the value $T_c \leq 0.1\tilde{\Omega}$ (see Eq.(1.1), $\tilde{\Omega} \approx \Omega_D$) provides the upper limit of the critical temperature. At the same time, as we know, there exist many superconductors with $\lambda > 0.5$  (e.g., Sn, Pb, Hg ). To clarify this point, an analysis, based on rigorous adiabatic theory   was carried out  ( Brovman and Kagan, 1967, Geilikman ,1971). It has been shown that the use of the experimentally observable acoustic law for $\Omega_0(q)$ is not self-consistent. The thing is that the electron-ion interaction participates in formation of the phonon spectrum in the system, so that one is dealing with a double counting.

Starting from the adiabatic theory with the Hamiltonian $\hat{H} = \hat{T}_{\vec{r}} + \hat{T}_{\vec{R}} + V(\vec{r},\vec{R})$ ($\hat{T}_{\vec{r}}$ and $\hat{T}_{\vec{R}}$ are kinetic energy operators for electrons and ions, and $V(\vec{r},\vec{R})$ is the sum of the Coulomb interactions) one can evaluate the electron-phonon interaction and the phonon spectrum rigorously. The electron-phonon interaction, indeed, leads to the formation of the experimentally observed phonon spectrum, and the aforementioned limitation on value of the coupling constant is absent.

There are superconductors with large values of the coupling constant (e.g., λ≈2.6 for Am-$Pb_{0.45}Bi_{0.55}$, see Wolf, 2012). The combination of high characteristic phonon frequency and large coupling constant can provide the high temperature superconducting state. This is the case for the sulfur hydrides, where such combination leads to high value of $T_c$ (see below, Ch. IV)

Note also, that the complex structure of the phonon spectra requires a modification of the usual methods, as it will be discussed below (Sec.IVB).

**B. Function $\alpha^2(\Omega)F(\Omega)$. Tunneling spectroscopy.**

Tunneling spectroscopy of ordinary metals is the uniquely powerful tool allowing us to obtain an important information about the energy spectrum of a superconductor. To be more concrete, with use of this technique, one can measure the value of the energy gap ,which includes the case of the multi-gap structure of the spectrum (the case is relevant to the hydrides, see Sec. VID ). Moreover, it allows one to evaluate such important quantity as the function $\alpha^2(\Omega)F(\Omega)$ , see above, Eqs.(3.1)-(3.4)

The tunneling contact contains two electrodes, separated by a barrier. For the most interesting case of the *S-I-N* system (*S* stands for a superconductor, *N* for a normal metal, and *I* for an insulator),one can obtain the following relation(Schrieffer et al., 1963;see also review by Scalapino,1969)

$$\sigma_s/\sigma_N = |\omega|\left[|\omega|^2 - \Delta^2(\omega)\right]^{-1/2} \quad (3.11)$$

Here $\sigma_s$ is the tunneling conductivity; $\sigma_s=\partial j/\partial V$, j is the tunneling current, and V is the applied voltage;$\Delta(\omega)$ corresponds to the analytical continuation of $\Delta(\omega_n)$ to real axis ; $\sigma_N$ is the conductivity for the *N-I-N* junction .The value of the energy gap $\varepsilon_0$ is determined by the relation $\omega = \Delta(i\omega)$.

The quantity $\sigma_s/\sigma_n$ can be measured experimentally (see Eq.(3.11). Since this quantity has a sharp peak at $\omega=\varepsilon_0$, the tunneling can be used for measuring the value of the energy gap that corresponds to the peak in the density of states. Note that the observation, say, of two peaks would manifest the presence of the two energy gaps ( see below, Sec.VI D ).

The special inversion procedure, allowing to reconstruct the function $\alpha^2(\Omega)F(\Omega)$ and the value of µ* was developed by McMillan and Rowell (see, e.g., Rowell,1969; Wolf, 2012). Usually the function $\alpha^2(\Omega)F(\Omega)$ contains peaks ; they correspond to peaks in phonon density of states . Note that, in turn, the function F (Ω) can be measured independently by neutron scattering technique. The coincidence of the peaks obtained by these two methods (i.e., tunneling spectroscopy and neutron scattering) is a crucial evidence of the fact that the pairing, indeed, is caused by the phonon mechanism.

For the sulfur hydrides the tunneling spectroscopy has not been performed yet. It would be interesting to develop tunneling spectroscopy and determine the important function $\alpha^2(\Omega)F(\Omega)$ for these new materials.

Note that the tunneling measurements under pressure were performed by Zavaritskii et al.,1971, to study properties of Pb . As mentioned above, this method has not been used so far for sulfur hydride and the function $\alpha^2(\Omega)F(\Omega)$ and the energy gap have not been determined experimentally. In what follows we are using the results of several theoretical papers describing the calculations of $\alpha^2(\Omega)F(\Omega)$ performed with use of the density functional formalism (see below, Sec.IVA).

## V. Sulfur hydrides.

### A. Phonon spectrum and the electron-phonon interaction

Let us now turn our attention to the material of interest, sulfur hydride. As was noted above, the Cooper pairing in the superconducting state is provided by the electron-phonon interaction and the main role is played by high frequency optical modes; this mechanism is manifested in the large value of the isotope coefficient for substitution of deuterium for hydrogen (see below, Ch. )

In principle, the value of the critical temperature can be evaluated with help of Eqs. (3.1),(3.3),which contain the function $\alpha^2(\Omega)F(\Omega)$. As mentioned above, the tunneling measurements allowing to reconstruct this function have not been performed yet. Instead ,we use results of the calculations carried out by several groups. For example, Fig. 6 shows the function $\alpha^2(\Omega)F(\Omega)$ calculated by Duan et al.(2014) for both, the high $T_c$ phase ($T_c$=203K) and for the structure with the lower value of the critical temperature ($T_c \approx$120K). For comparison, one can see the pictures for Pb, the conventional superconductor (Fig.5). The lattice dynamics and superconducting properties were treated with use of the density functional theory (see, e.g.,Baroni et al.,2001) and the quantum expresso code (Paolo,2009). Results for the function $\alpha^2(\Omega)F(\Omega)$ are also given in Errea et al.,2015;Sono et al.,2016;Durajsky et al.,2015. As was mentioned above, all these calculations are based on the density functional formalism, but ,because of using different codes, they differ in details, which lead to some spread in the numerical results. Nevertheless, the calculated density of states , $F(\Omega)$ ,and the functions $\alpha^2(\Omega)F(\Omega)$ are rather similar.

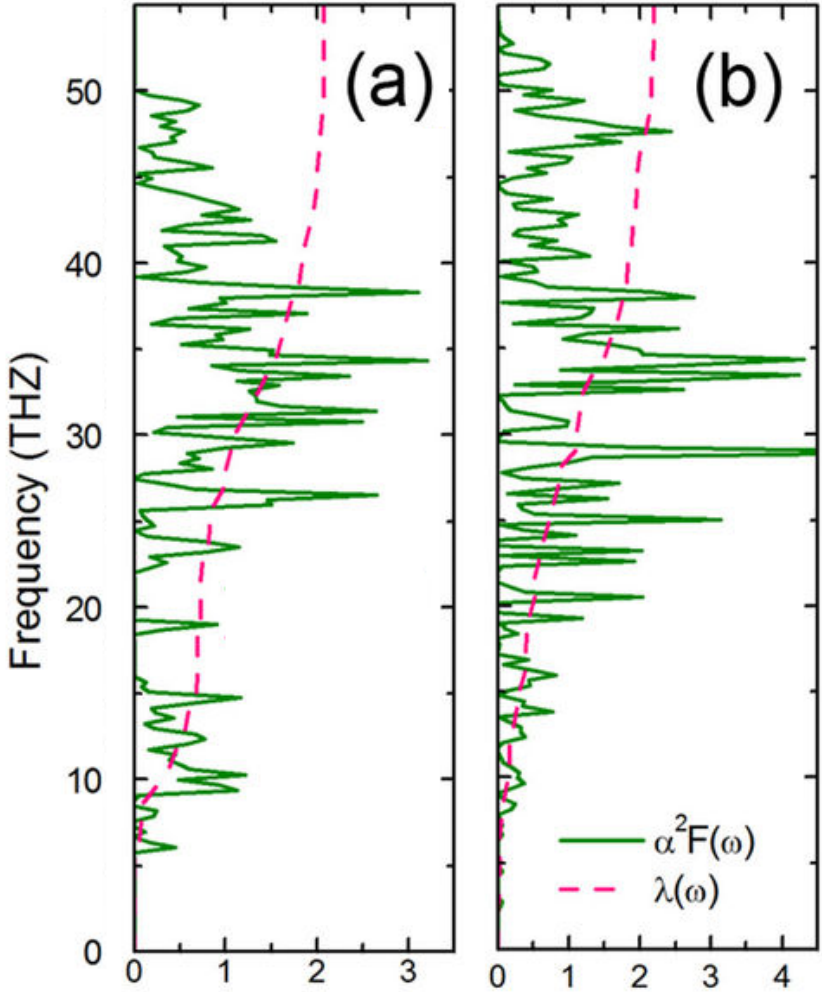

Fig. 6. The spectral function $\alpha^2(\Omega)F(\Omega)$ and the electron-phonon coupling constant $\lambda(\Omega)$ (dash lines) for the "low" $T_c$ and high $T_c$ phases of the sulfur hydride. The function $\lambda(\Omega)$ is defined by Eq.(3.4) with the phonon frequency $\Omega$ as the upper limit : (a) the functions $\alpha^2(\Omega)F(\Omega)$ and $\lambda(\Omega)$ for the "low" $T_c$ phase; (b) the functions $\alpha^2(\Omega)F(\Omega)$ and $\lambda(\Omega)$ for the high $T_c$ phase. Units: dimensionless quantities $\alpha^2(\Omega)F(\Omega)$ and $\lambda(\Omega)$ (horizontal axis) and the phonon frequency ( vertical axis). The acoustic and optical branches are separated at ~15THZ .From Duan et al.,2014.

One can see from Fig. 6 that for sulfur hydride the function $\alpha^2(\Omega)F(\Omega)$ has a rather complicated structure. Indeed, in addition to acoustic modes, it contains a number of optical phonons. Moreover, because of high frequencies of optical modes, the phonon spectrum is broad and extended up to 200 meV (Fig.5c).

As mentioned above (Sec. IIIA), the value of the critical temperature can be found without invoking the coupling constant concept (Errea et al.,2015;Akashi et al.,2015;Flores-Livas et al.,2016;Sano et al.,2016). Such a program was carried out in the framework of the superconducting density functional theory (see, e.g., Luders et al.,2005). Errea et al. ,2015, used the equation, which looks like Eq.(3.1) with the replacement: $|\omega_n| \mapsto [\omega_n^2 + \Delta^2(\omega_n)]^{1/2}$ in the denominator of the integrand. This equation is valid at any temperature. The value of $T_c$ was calculated from such a non-linear equation by successive approximations. With each iteration the value of order parameter decreases, and $T_c$ was identified as the point (after up to 512 iterations) where the order parameter approaches zero.

According to Errea et al.(2015 ) and Sano et al.(2016), an important role is played by anharmonicity. For example, taking anharmonicity into account noticeably shifts the value of $T_c$ ( by as much as up to ~ 60K : from $T_c \approx 260K$ to $T_c \approx 200K$ in the high $T_c$ phase. The role of anharmonicity and quantum effects was analyzed with the use of the so-called stochastic self-consistent harmonic approximation (SSCHA), developed by Errea et al. (2014). The impact of the zero point motion was analyzed by Bianconi and Jarlborg (2015a) and by Sano et al. (2016). The effects of anharmonicity and zero point motion could be essential, because of the small mass of the hydrogen ions. The problems raised in the aforementioned papers deserve a further study.

As was noticed in the previous section, the concept of the coupling constant was very fruitful for studying conventional superconductors. Similar approach was used by Li et al.(2014),by Papaconstanopoulos et al.(2015), to calculate $T_c$ for sulfur hydrides. Since the value of the total coupling constant is large $\lambda \gtrsim 2$ (see Fig. 6), they did not use Eq.(3.8),but more general expression (Allen and Dynes,1975):

$$T_c = \frac{f_1 f_2 \tilde{\Omega}_{\log}}{1.2} \exp\left[-\frac{1.04(1+\lambda)}{\lambda - \mu^* - 0.62\lambda\mu^*}\right]$$

(4.1)

Eq.(4.1) is similar to Eq.(3.8), but the pre-exponential factor is multiplied by the product $f_1 f_2$; the functions $f_1$ and $f_2$ are numerically fit for the solution valid at larger $\lambda$ and have a form:

$$f_1 = \left[1 + (\lambda/\Lambda_1)^{3/2}\right]^{1/3}; f_2 = 1 + \lambda^2 \left(\tilde{\Omega}/\Omega_{\log} - 1\right)\left(\lambda^2 + \Lambda_2^2\right)^{-1}$$

$$\Lambda_1 = 2.46(1 + 3.8\mu^*); \Lambda_2 = 1.82(1 + 6.3\mu^*)\left(\tilde{\Omega}/\Omega_{\log}\right)$$

The large value of the constant for the coupling to characteristic phonon frequency leads to a high $T_c$ close to that observed experimentally.

The phonon spectrum of the sulfur hydrides is broad and rather complex: it contains a number of optical and acoustic branches. On the other hand, these branches are well separated, and this separation allows us to develop different approach, which will be described in the next section.

**B. Generalized equation and two coupling constants.**

In the more general approach (Gor'kov and Kresin, 2016) the value of $T_c$ was evaluated analytically. As will be shown, it allows us to compare the relative contributions of the optical and acoustic modes for different parts of the phase diagram.

The phonon spectrum contains two well-separated frequency intervals for the optical and acoustic branches. Let us introduce the coupling constants $\lambda_{opt.}$ and $\lambda_{ac}$ for each of these regions and the corresponding average frequencies $\tilde{\Omega}_{opt}, \tilde{\Omega}_{ac}$ and .Then the equation for the order parameter at $T=T_c$ takes the following form (cf. Eq. (3.5)):

$$\Delta(\omega_n) = \pi T_c \sum_m \left[ (\lambda_{opt} - \mu^*) D(\tilde{\Omega}_{opt}, \omega_n - \omega_m) + \lambda_{ac} D(\tilde{\Omega}_{ac}, \omega_n - \omega_m) \right] \frac{\Delta(\omega_m)}{|\omega_m|}$$

(4.2)

Here

$$\lambda_i = \int_i d\Omega \alpha^2(\Omega) F(\Omega)/\Omega ; \Omega_i = <\Omega^2>_i^{1/2}$$

(4.2')

$$<\Omega^2>_i = (2/\lambda) \int_i d\Omega \Omega \alpha^2(\Omega) F(\Omega) \ ; \ i \equiv \{opt., ac.\}.$$

The critical temperature must be calculated with the use of Eq. (4.2). This equation is the generalization of Eq.(3.5) for the presence of two phonon groups, the acoustic and optical modes.

### B. Critical temperatures for different phases.

Let us assume that in the high-$T_c$ phase (Im-3m) $\lambda_{opt.} \gg \lambda_{ac}$ We also suppose that $\lambda_{opt} \tilde{<} 1.5$ . As will be shown below, these conditions are indeed satisfied. Then from Eqs. (4.2) and (4.2') one can obtain expression for $T_c$ in an analytical form and, hence, evaluate the value of the critical temperature for the sulfur hydrides.

Let us write $T_c$ as

$$T_c = T_c^0 + \Delta T_c^{ac}; T_c^0 \equiv T_c^{opt} \qquad (4.3)$$

and assume that $\Delta T_c^{ac} \ll T_c^0$. Based on Eq.(4.2), with use of these assumptions, one can obtain the following analytical expression for the critical temperature ion the high $T_c$ phase:(Gor'kov and Kresin,2016):

$$T_c = \left[1 + 2\frac{\lambda_{ac}}{\lambda_{opt} - \mu^*}\frac{1}{1+\left(\pi T_c^0/\Omega_{ac}\right)^2}\right]T_c^0 \qquad (4.4)$$

For $T_c^0$, one can use the McMillan-Dynes expression (Eq.(3.8)), which is valid for $\lambda_{opt} \lesssim 1.5$:

$$T_c^0 = (\tilde{\Omega}_{opt}/1.2)\exp\left[-\frac{1.04(1+\lambda_{opt})}{\lambda_{opt} - \mu^*(1+0.62\lambda_{opt})}\right] \qquad (4.5)$$

One can use the close expression, obtained analytically by Geilikman et al.(1975), valid also for $\lambda_{opt} \gtrsim 1.5$:

$$T_c^0 \approx \tilde{\Omega}_{opt} \exp\left[-\frac{1+1.5\lambda_{opt}}{\lambda_{opt} - \mu^*(1+0.5\lambda_{opt})}\right] \qquad (4.6)$$

As is known that the coefficients in the Eq.(4.5) were selected to fit the data for Nb. As for Eq.(4.6), it was obtained by analytical solution of the Eq. (3.5). Note that these expressions are rather similar. Indeed, by neglecting $\mu^*$ for simplicity, one can write Eqs.(4.5),(4.6) in the form: $T_c \approx A\exp(-1/\lambda_{opt})$, with close values of the pre-exponential factor.

For ordinary superconductors the values of the coupling constants and $\mu^*$ (usually $\mu^* \approx 0.1 \div 0.15$) can be determined from tunneling spectroscopy measurements (see, e.g., Wolf, 2012). For sulfur hydride, we deduce the coupling constants $\lambda_{opt}$ and $\lambda_{ac}$ from several theoretical calculations of $\alpha^2(\Omega)F(\Omega)$. Although the corresponding theoretical results differ somewhat, they are relatively close. The values of $\lambda_{opt}$ and $\lambda_{ac}$ can be directly determined from $\lambda(\Omega)$ (Fig.6). We estimate $\lambda_{opt} \approx 1.5$ and $\lambda_{ac} \approx 0.5$; see Fig. 6b. We assumed (see above) that $\lambda_{opt} \gtrsim 1.5$ and $\lambda_{opt} \gg \lambda_{ac}$. One can see that the obtained values are consistent with the above approximations. Using these coupling constants and taking the values $\tilde{\Omega}_{opt} = 1700K$ and $\tilde{\Omega}_{ac} = 450K$ ( $\mu^* \approx 0.14$, which is close to that for usual superconductors and was also calculated by Flores-Livas et al.,2016), we obtain $T_c^0 \equiv T_c^{opt}$ =170K and $\Delta T_c^{ac} = 45K$ , so that in total $T_c \approx 215$ K, in quite good agreement with the value of $T_c \approx 203$ K observed by Drozdov, Eremets et al. ( 2015). The main contribution comes from the optical phonons; this confirms the self-consistency of our approach.

The fact that the coupling constant $\lambda_{opt}$ in the *cubic* phase is so large is the key factor underlying the observed high $T_c \approx 203$ K. Qualitatively, this comes about due to the ability of sulfur to retain several hydrogen atoms in its proximity, that is, to the presence of many light ligands near the *S* atoms. There are six (!) ligands in the high $T_c$ phase (see Fig. 3c ).

To demonstrate the importance of this point, let us evaluate the value of $T_c$ for the "low $T_c$" phase (R3m structure). One can see from Fig. 6a (Duan et al.,2014) that the coupling constants for this phase are: $\lambda_{opt} \approx \lambda_{ac} \approx 1$. It is interesting that in this case the value of the total coupling constant $\lambda_T = \lambda_{opt} + \lambda_{ac} \approx 2$ and is close to that in the high $T_c$ phase (see above).However, the relative contributions are shifted towards to low frequencies; the value of $\lambda_{ac}$ is larger for the "low $T_c$" phase. In this case $T_c < \tilde{\Omega}_{ac} << \tilde{\Omega}_{opt}$ , and one can estimate $T_c$ within the usual BCS logarithmic approximation while adding the renormalization function Z≈1+ $\lambda_T$ into the exponent (Grimvall,1981).:

$$T_c \approx \left(\tilde{\Omega}_{opt}\right)^{\frac{\lambda_{opt}}{\lambda_T}} \left(\tilde{\Omega}_{ac}\right)^{\frac{\lambda_{ac}}{\lambda_T}} \exp\left[-\frac{1+\lambda_T}{\lambda_T - \mu^*}\right] \qquad (4.7)$$

With $\tilde{\Omega}_{opt} \approx 105 meV$, $\tilde{\Omega}_{ac} \approx 26 meV$ (Duan et al.,2014),we obtain $T_c \approx 120K$.

The transition into the high $T_c$ phase is accompanied by re-distribution of interaction of electrons with optical phonons and their interaction with acoustic branches. This re-distribution is manifested in an increase in a number of hydrogen ligands, caused by the structural transition. This is a key factor determining record high value of $T_c$.

Dividing the phonon spectrum and, correspondingly, the electron-phonon interaction in two parts, turns out to be rather fruitful. First of all, the value of $\lambda_{opt}$ is within the range of applicability of Eq. (4.4). Moreover, one can evaluate the relative contribution of the optical and acoustic branches of the phonon spectrum to $T_c$. For the high $T_c$ phase the contribution from the optical phonons comprises ~80% and only ~20% is due to the acoustic part. The impact of acoustic phonons is noticeably smaller than that of optical branches (45K vs. 170K), but still is essential.

The method proposed above can be of relevance for other materials as well. A promising example is calcium hydride, $CaH_6$ (see below, Sec. VIIA ).

It is known that for a number of superconductors the value of the electron-phonon coupling constant is large. Among them Pb

($\lambda\approx$1.55), Hg ($\lambda\approx$1.6), AmPb$_{0.45}$Bi$_{0.55}$ ($\lambda\approx$2.6), see, e.g., Wolf, 2012. However, because of low values of characteristic phonon frequencies, the values of the critical temperature for them are not large. An uniqueness of the hydrides is that they combine the strong coupling, especially to optical modes, with high values of the characteristic frequencies.

**V. Isotope effect**

According to Drozdov et al.(2015), the substitution of deuterium for hydrogen noticeably affect the value of the critical temperature. Observation of this isotope effect is of fundamental importance, since it proves (a) that the high $T_c$ state is caused by the electron-phonon interaction, and (b) that namely the high frequencies hydrogen modes determine the value of $T_c$. Indeed, the optical modes are mainly due to motion of hydrogen, whereas for the acoustic modes the participation of sulfur ions prevails. Therefore the magnitude of the isotope shift for the deuterium for hydrogen substitution reflects indirectly the relative contributions of the each group (the optical *vs.* acoustic ones) into the observed $T_c$.

Defining the isotope coefficient via the relation $T_c \propto M^{-\alpha}$ one can obtain the following expression for $\alpha$ (see,e.g.,Gor'kov and Kresion,2016):

$$\alpha = -(M/T_c)(\partial T_c / \partial \tilde{\Omega})(\partial \tilde{\Omega} / \partial M) \quad (5.1)$$

Since the deuterium substitution affects the optical modes, one can write the expression (5.1) in the form ( in the harmonic approximation; then $\tilde{\Omega} \propto M^{-1/2}$ ):

$$\alpha = 0.5\left(\tilde{\Omega}_{opt}/T_c\right)\left(\partial T/\tilde{\Omega}_{opt}\right) \quad (5.2)$$

The value of the isotope coefficient in the high $T_c$ phase can be calculated with use of Eqs. (4.4) and (5.2). As a result, we obtain:

$$\alpha \approx \frac{1}{2}\left[1 - 4\frac{\lambda_{ac}}{\lambda_{opt}}\frac{\rho^2}{\left(\rho^2+1\right)^2}\right] \quad (5.3)$$

Here $\rho = \tilde{\Omega}_{ac}/\pi T_c^0$. With $\lambda_{opt} \approx 1.5, \lambda_{ac} \approx 0.5, \tilde{\Omega}_{opt} \approx 450K$, we obtain $\alpha \approx 0.35$, in a good agreement with the experimental data in Fig.4.

One should use the different expression, Eq.(4.7), for $T_c$ in the "low $T_c$" phase. Then, with use of Eq.(5.2), one can find $\alpha \approx 0.25$, which is noticeably smaller than that for the high $T_c$ phase. Note that the agreement between the value obtained from Eq.(5.3) and the experimental data is rather good for the high $T_c$ phase. As for the "low" $T_c$ phase, the data are not so well determined. It would be interesting to perform the measurements at lower pressures, at the values more distant from the region of the transition into the high $T_c$ phase.

As was mentioned above (Sec. IIB), at low pressures a formation of the $H_3S$ phase occurs (see Eq. (2.1)). According to Akashi et al.

(2016), the transformation occurs through intermediate structures. In other words, we are dealing with the coexistence of phases and the percolation scenario, so the percolation threshold corresponds to formation of the so-called infinite cluster, that is, to the metallic state. As a result and the *R3m* phase contains some inclusions. Then one can expect the pressure dependence of the isotope coefficient inside of the R3m phase.

Note that the usual analytical derivation of the value of the isotope coefficient is carried out under the assumption that the ionic mass is the continues variable. It is essential that the obtained value of $\alpha$ appears to be independent of M. It allows to use it for different values of the ionic mass, even the latest has discrete values. Qualitatively this means that the shift in the value of $T_c$ stays the same within the phase, so that the dependences $T_c(P)$ for different isotopes are parallel to each other. For example, in our case the shift is described by the relation : $T_c^H / T_c^D = (M_D/M_H)$. In the high $T_c$ phase $\alpha \approx 0.35$ (see above), and, therefore, $T_c^H / T_c^D = 1.25$. Since $T_c^H = 203K$, we obtain $T_c^D \approx 165$ K. This value is in a rather good agreement with the measurements by Drozdov et al.(2015)

The value of the isotope coefficient in the high $T_c$ phase is relatively large and it reflects the fact that the pairing in this phase is dominated by the optical H-modes, whereas in the "low $T_c$" phase the contributions of the optical and acoustic modes are comparable. The impact of the isotopic substitution in the region of smaller $T_c$ is weaker than in the high $T_c$ phase. A smaller $\alpha$ is in agreement with the larger role played by the optical phonons in the *cubic* high $T_c$ phase

Notice that the value of $\alpha$ can be affected by the anharmonicity (Errea et al.,2015) and by the dependence of $\mu^*$ on $\tilde{\Omega}_{opt}$, although the last contribution is of the order of $(\mu^*/\lambda_{opt})^2$ and is small. However, the main conclusion that the value of the isotope coefficient depends on pressure and is different in different phases, remains valid and reflects the relative contributions of the optical and acoustic modes.

**VI. Energy spectrum of the high $T_c$ phase: two-gap structure and non-monotonic dependence of $T_c$.**

**A. Structural transition. High $T_c$ phase.**

As was mentioned above, the pressure dependence of    is highly asymmetric relative to its maximum value     $T_{c;max}$ =203K. The value of $T_c \approx$ 120K at $P \approx$ 125GPa sharply increases to $T_c \approx$ 200K at $P \approx$ 150GPa (Drozdov et al.,2015; Einaga et al.,2016;) .A rapid increase in $T_c$ is followed by a slow decrease at $P>P_{cr.} \approx$ 150GPa (Fig.4). Remarkably, the structural transition into the *high $T_c$* phase takes place somewhere in the same pressure interval where it is accompanied by a sharp increase in the value of the critical temperature. Currently, it is generally accepted that the Bravais lattices of the high- $T_c$ and the low- $T_c$ phases belong to different symmetries (the *Im-3m* group for the high $T_c$ cubic phase and the trigonal *R3m* symmetry group for the low-$T_c$ phase) and that the structural phase transition between them occurs at a pressure somewhere in between $P \approx 150 GPa$ and $P \approx 125 GPa$ . The sharpness of the increase prompts the question whether transition into the high- *Tc* phase could be the first order transition. This scenario was discussed by us (Gor'kov and Kresin,2017) with use of the group theory and taking into consideration the impact of lattice deformations. The picture is similar to those considered by Larkin and Pikin (1969) and later by Borzykin and Gor'kov (2009): the coupling to lattice can transform the second order transition into the transition of the first order. The idea of the first order transition allows us to explain self-consistently the slow decrease in Tc with an increase in pressure above the pressure $P_{cr}$ corresponding to the maximum of $T_c$ $\approx 203K$ . The appearance of the two-gap spectrum is an important ingredient of the picture (see below, Sec VIC)

According to the band structure calculations (Duan et al.,2014; Flores-Levas et al.,2016; Errea et al.,2015,2016; Akashi et al.,2015;Neil and Boeri,2015), the high $T_c$ compound is characterized by broad energy bands ( large Fermi surface) and strong interaction between electrons and high frequency optical phonons. The calculated values of $T_c$ and the isotope coefficient are in a good agreement with the experimental data (see above, Sec.IVC, Ch.V ).

Meanwhile, the calculations also revealed the presence in the high-$T_C$ phase of small Fermi-pockets. The importance of their existence was emphasized by Bianconi and Jarlborg (2015[a,b,c]),who suggested that the pockets play an especial role in increasing $T_c$. Note, however, that the analysis of the electron-phonon mediated pairing on pockets should be carried out with a considerable care. Let us discuss this point in more details.

### B. Migdal adiabaticity criterion and small pockets.

The complex structure of the Fermi surface with small pockets emerging in addition to several large Fermi sheets is not uncommon for many novel superconductors such, for instance, as the high-$T_c$ oxides, low-dimensional organic conductors, the so-called heavy fermions (see, e.g., Gor'kov,2012). Here we focus on the possible impact of small pockets on the superconductivity in hydrides.

The main equation, Eq. (3.1) is valid if the so-called adiabatic parameter $\left(\tilde{\Omega}/E_F\right)$ is small ($\left(\tilde{\Omega}/E_F\right)$<<1 ;Migdal,1958, see also review by Scalapino, 1969). Then one can neglect all higher order corrections (so called "vertex corrections"),containing the products of the matrix elements of the electron-phonon interaction. Then the right side of equation for the order parameter is linear in the coupling constant λ (see Eq.(3.1) and also Eq.(3.3)).The value of the coupling constant is expressed by Eq. (3.10). According to Migdal (1958), the correction to Eq.(3.1) contains an additional term $\propto \lambda^3 \left( \tilde{\Omega} / E_F \right)$.

Therefore, the inequality $\tilde{\Omega} \ll E_F$ allows us neglect the higher order corrections. If $\Omega \gtrsim E_F$ one should include the contribution of all higher terms. But the rigorous calculations of even the second term (Grimaldi et al.,1995) appear to be the non-trivial task. At this point we meet with the problem, which at the present time remains unresolved.

The case of the weak electron-phonon coupling ($\lambda \ll 1$) is the exception (see Gor'kov, 2016). Because of the smallness of $\lambda$, the vortex corrections can be neglected.

The condition $\left( \tilde{\Omega} / E_F \right) \ll 1$ is satisfied for most conventional superconductors, since in usual metals the Fermi energy is large compared with the Debay energy $E_F \gg \tilde{\Omega} \approx \Omega_D$. However, this is not the case for pockets in the sulfur hydrides, since for these materials the characteristic frequency of optical modes $\tilde{\Omega}_{opt} \approx (1.5-2)100 meV$, whereas the Fermi energy of a pocket is of order $\approx$ (40-50) meV. The electron-phonon interaction on the pockets can be rigorously treated only in the case of weak coupling, that is, if the corresponding coupling constant $\lambda_P \ll 1$.

### C. Broad bands and pockets.

The calculated spectrum of electrons display small pockets only inside of the high- $T_c$ phase and it may be tempting to relate the high value of $T_c$ to the appearance of the pockets (Bianconi and Jarlborg, 2015[a,b,c]; Quan and Pickett, 2016). In this scenario, the major pairing interaction occurs on the pockets. As for the electron-phonon interaction on the larger bands, it is weak and is playing only a secondary role.

At this point it is worth noting that the calculations of $T_c$ performed assuming the prevailing role of the large bands and sufficiently strong coupling are in a good agreement with the experimental data (see Ch. IV), so that there is no special need for modifying the picture. Besides of that, if one is trying to assign the leading role to pockets, then it is clear that the on- pocket interactions should be rather strong in order to provide high $T_c$ . However, in this case the rigorous treatment is not known, because of the violation of the Migdal theorem (see above). On the other hand, if the main contributions into the interaction were coming from the large bands, then the contribution of the pockets could be assumed weak, and the case can be analyzed self-consistently.

Note also-and this is a strong argument- that if the leading role of pockets were due to a peak in their density of states ,this would produce a pre-factor in the expression for $T_c$ of an electronic origin. However, such a pre-factor cannot depend on the ionic mass, in the strong contradiction with the observed isotope effect (Drozdov et al., 2015).

**D. Two-gap spectrum. Slow decrease in $T_c$ at $P>P_{cr}$**

In the superconducting state the pockets are characterized by the energy gaps in their electronic spectrum. Below we consider such two-gap model with one gap corresponding to the broad band and the second gap describing the excitations on the pocket.

The two-gap model was introduced shortly after the creation of the BCS theory (Suhl et al.,1959; Moskalenko,1959). From that follows, we stipulate that under the notion of the two gap spectrum we mean the presence of two peaks in the density of states.

Each band has its own set of the Cooper pairs. Since a single pair is formed by two electrons with equal and opposite momenta, one can neglect pairing between electrons belonging to different bands. Indeed, in general, the electrons on the Fermi level, which belong to different bands, have different values of the momenta. However, in the two-band model, the absence of the interband pairing does not mean that the pairing within each band is insensitive to the presence of the other band. Indeed, the presence of the second band gives rise to an additional pairing channel. Namely, the electron originally located on the first band can radiate phonon and make the virtual transition into the second band. The second electron can absorb the phonon and also make transition into second band forming the pair the first electron. Therefore, owing to the interband electron-phonon scattering the on-the-pocket electrons can form the Cooper pairs on the large Fermi surface and vice versa.

Let us stress one important point. As noted above, the two gap model was introduced shortly after the creation of the BCS theory. Nevertheless, the two gap phenomenon has not been essential for the conventional superconductors. This is due to the large coherence length; more specifically, the inequality l<< $\xi$ ( l is the mean free path) ,which hold for usual superconductors , leads to the averaging caused by the interband impurity scattering. As a result ,the two gap picture is washed out and the usual one gap picture is applicable.

The two-gap spectrum was observed for the first time in the *Nb*-doped *SrTiO$_3$* system (Binnig et al,1980) with use of the scanning microscope (STM) technique. The second gap appears as a result of doping and filling the second gap. At present, the two-gap picture is important feature of the novel superconducting systems, and this is due to their short coherence length. It has been observed in the cuprates (Greek et al.,1988),in MgB$_2$ (Uchiyama et al., 2002 ; Tsuda et al.,2003), see review by Kresin et al.(2014).

At the formulation of the two-gap model for the high- $T_c$ phase of sulfur hydrides one can introduce three coupling constants: $\lambda_L$ , responsible for strong electron-phonon interactions on the large band , $\lambda_P$ <<1 ( weak coupling on the pockets), and $\lambda_{LP}$ <<1, describing the transitions from large band electrons to the pairing states on the pockets. The coupling constants $\lambda_L$, $\lambda_P$ and $\lambda_{LP}$ are described by Eq.(3.10);the constant $\lambda_{LP}$ contains the matrix element describing the interband transitions caused by the electron-phonon interaction. Note also that, because of the interband transitions, the system has the common temperature of the superconducting transition $T_c$. In addition, their presence is beneficial for superconductivity.

Performing the calculations (Gor'kov and Kresin,2016),one can show that the shift in $T_c$ caused by the presence of pockets is proportional to the density of the states on the pockets: $\Delta T_c \propto \nu_F$

Return to the problem of the strong asymmetry in the pressure dependence of $T_c$ relative to the position of its maximum value (at $T_{c;max}$ =203K ( at $P_{cr.} \approx$150GPa) posed in Sec.IIC. Assume, as above, that the sharp increase in $T_c$ (from $T_c \approx$120K to $T_{c;max} \approx$ 200K ) is the result of the first order structural transition into high $T_C$ cubic phase. This phase is characterized by the coexistence of broad band (responsible for the large part of the Fermi surface) and small pockets.

As mentioned above, the interaction between the large band and the pocket leads to the shift in the temperature of transition $\Delta T_c = T_c - T_{c;0}$ ,which is proportional to the density of states on the pocket $\nu_P(E_F) \propto m_P P_{F;poc}$ where $m_P$ and $P_{F;poc}$ are the effective mass and momentum for the pocket's states, $T_{c;0}$ is the value of the critical temperature in the absence of the pockets. It is essential whether the pockets appear instantly, as a result of the discontinuous first order transition or the transition into the high- $T_c$ phase is either of the second order or is of a topological nature at which the pocket's size would grow continuously with the further increase in pressure.

Above we have given arguments in favor of the first order transition, which is accompanied by the emerging singularity in the density of states in the form of pockets. The further increase in pressure leads to shrinking of the pocket with an effective decrease in their Fermi momentum $P_{F;poc}$ and the corresponding depression of the two-gap picture. Since the two-gap scenario is beneficial for superconductivity, such a depression leads to a decrease in *Tc* .This explains the observed slow decrease in $T_C$ after the transition; the small scale of the decrease in $T_C$ at *P>P$_{cr}$* is related to small values of $\lambda_P$ and $\lambda_{PL}$

The two-gap spectrum and its evolution with pressure, including the decrease in the amplitude of the second gap at *P>P$_{cr}$* must be confirmed by future tunneling experiments. The presence of the second energy gap will be manifested as the second peak in the density of states.

## VII. Other hydrides

S-H system appears to be the first hydride, which displays record high value of *T$_c$.* We described above (Ch.II) the development of the field. Let us discuss here several other studied hydrides, which display interesting and promising properties.

As was emphasized in the Introduction and Ch.IV, the theoretical studies containing predictions of specific hydrides along with values of $T_c$ and the pressures, deserve a special credit. In this Chapter we will be talking about promising compounds. As was mentioned above, the group of Y.Ma made a very important prediction related to sulfur hydrides (Li et al,2014). This group made several other interesting predictions, which are waiting their future confirmations (Peng et al.,2017;see the review by Zhang et al.,2017). They predicted the high $T_c$ superconducting state for calcium hydride ($T_c \approx$240K; Wang et al,2014) and for $YH_6$ ($T_c \approx$264K; Li et al.,2015).Note that the compound $MgH_6$ ,similar to $CaH_6$ was also studied (Feng et al.,2015) and the values of $Tc \approx 260K$ at P $\gtrsim 300GPa$ were predicted. Even higher values of $T_c$ were obtained by Szczesniak and Durajski (2016). The next section contains the description of the calcium hydride and $MgH_6$; these are examples of the future high $T_c$ compounds.

### A. Calcium hydride; $MgH_6$

Calcium hydride, $CaH_6$, has been analyzed by Wang et al. (2012) and looks very promising. Wang et al. evaluated its structure and the function $\alpha^2(\Omega)F(\Omega)$ (Fig.7 ).Based on numerical solution of Eqs.( 3.1),(3.3), they predicted that at pressures *P≈150GPa* the value of $T_c$ will be higher than that for $H_3S$.

The value of $T_c$ for $CaH_6$ can be evaluated with use of Eq.(4.2). Indeed, in accordance with an approach described in Sec. IVB, the electron-phonon interaction can be separated in two parts. From Fig. 7 one can determine that $\lambda_{opt} \approx 2.1$, $\lambda_{ac} \approx 0.6$, $\tilde{\Omega}_{opt} = 820$ sm$^{-1}$, $\tilde{\Omega}_{ac} = 350$ sm$^{-1}$. Correspondingly, one can write that $T_c = T_c^0 + \Delta T_c^{ac}; T_c^0 \equiv T_c^{opt}$ is determined by contribution of the optical modes. However, because of large value of $\lambda_{opt}$, the MacMillan-Dynes expression for $T_c^0$, (Eq.(3.7), is not applicable (it is valid for $\lambda_{opt} \lesssim 1.5$). One can use the modified MacMillan equation, Eq. (4.1), valid for larger $\lambda$. Another option is to use the analytical expression, valid for any value of the coupling constant (Kresin,1984;see also review by Kresin and Wolf,2009):

$$T_c^0 = \frac{0.25 \tilde{\Omega}_{opt}}{\left[e^{2/\lambda_{eff}} - 1\right]^{1/2}} \quad (7.1)$$

$$\lambda_{eff} = (\lambda_{opt} - \mu^*)\left[1 + 2\mu^* + \lambda_{opt}\mu^* t(\lambda_{opt})\right]; t(x) = 1.5\exp(-0.28x)$$

With use of Eq. (7.1) and the parameters for $CaH_6$, we obtain $T_c \approx 230$K. This value contains the contributions of the optical ($T_c^0$) and acoustic ($\Delta T_c$) modes: $T_c^0 \approx 180$K, and $\Delta T_c \approx 50$K. Therefore, the optical and acoustic modes contribute 78% and 22% to the total value of the critical temperature, correspondingly. As for the isotope coefficient, one can obtain from Eq. (5.2) the value $\alpha \approx 0.36$; it is close to that for the high phase of the sulfur hydrides.

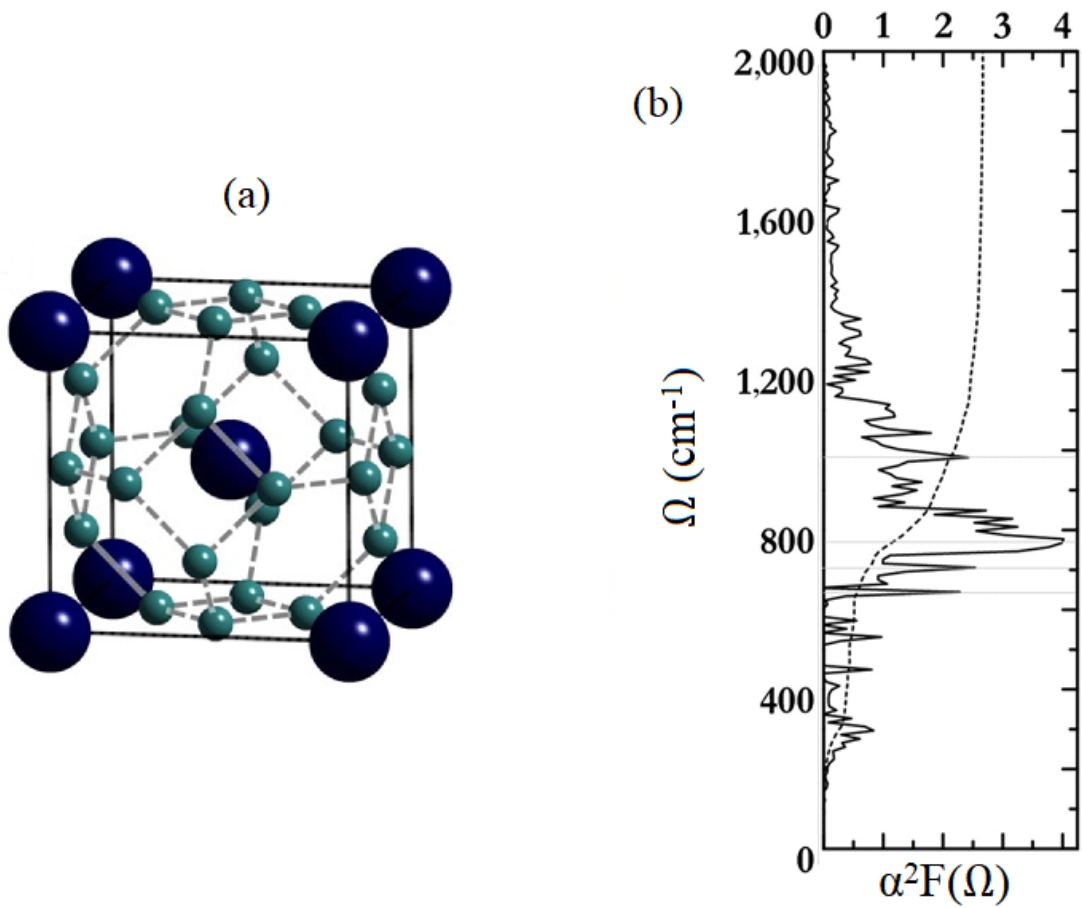

Fig. 7. Superconducting calcium hydride:(a) structure of CaH$_6$ ; (b) the spectral function $\alpha^2(\Omega)F(\Omega)$ for CaH$_6$. From Wang et al., 2012.

The hydride MgH$_6$ studied by Feng et al.,2015, and by Szczesniak and Durajski,2016 has properties similar to those for the calcium hydride. The structure contains the sodalite-like hydrogen cage with interstitial Ca (Mg) atoms (Fig.7). This is not occasional, since the Mg and Ca atoms have similar chemical properties. As mentioned above, the predicted values of $T_c$ are even higher than for sulfur hydrides.

**B. Palladium hydride.**

In principle, palladium hydride ( Pd-H), is not a new superconductor. It has been discovered in 1972 ( Stritzer and Buckel,1972 ); its $T_c \approx$8-10$K$. This hydride was known by the value of the isotope coefficient, which appears to be negative. This phenomenon was explained by strong anharmonicity (Ganzuly,1973; Klein and Cohen,1992), namely, by the peculiar dependence of the phonon frequency on ionic mass.

According to the study by Syed et al.,2016, one can observe the superconducting state of Pd-H at higher temperatures ($T_c \approx$54K for Pd-H and $T_c \approx$60K for Pd-D; the isotope coefficient is negative, as for the usual compound).It is important that such an increase in $T_c$ has been observed at the ambient pressure. It occurs, thanks to special sample preparation, namely, because of fast cooling of the hydride.

The result looks interesting, because the superconducting state persists up to high temperature at ambient pressure. The authors observed a drastic drop in resistance. However, the Meissner effect has not been demonstrated yet, as well as the impact of the magnetic field on $T_c$. Of course, it would be interesting if future experiments confirm the presence of superconductivity at elevated temperatures.

## C. Transition of ice under high pressure and by doping

In general, the transition of an insulator into metallic state can proceed through two channels: either via doping (e.g., high $T_c$ cuprates, fullerides), or by increase in pressure (e.g., hydrides). The recent paper (Flores-Livas et al., 2016) is interesting, because the authors consider theoretically the transition, caused by combination of these channels. They studied the properties of $H_2O$, which can be transferred into solid phase (ice) by applied pressure ( up to $P \approx 150GPa$). The crystal continues to be an insulator at this pressure. It has been proposed that the sample in the solid phase can be doped by nitrogen. The calculation demonstrates that, as a result of such a doping under the pressure, the material becomes metallic and even superconducting with rather high value of $T_c$ $\approx 60K$.

Note that at ambient pressure and also at low pressure (up to $P \approx 110GPa$) the oxygen ion has four hydrogen neighbors (so called phase I). Two of these neighbors are covalently bonded with oxygen and form the $H_2O$ molecule ,and other two ions formed additional hydrogen bonds. The lengths of the bonds are different and the structure is asymmetric. However at higher pressures (of the order of $P \approx 300GPa$) the so-called ice -X phase is formed and it is characterized by symmetric O-H bonds      ( see Goncharov et al., 1999)

As was noted above, at $P \approx 150GPa$ the ice crystal is still in the insulating state. To prompt a transition into metallic state one needs to use doping and it turns out that the nitrogen is the best dopant.

The nitrogen for the oxygen substitution leads to the hole conductivity. Moreover, the transition into a metallic state is accompanied by changes in the phonon spectrum. All these changes provide the transition into superconducting state. The calculations show that best value of the superconducting parameters correspond to relatively low doping (4-6%). The calculated $\alpha^2(\Omega)F(\Omega)$ function leads to the value of $T_c \approx 60K$. Of course, this value is below $T_c \approx 203K$ observed for $H_3S$ phase of sulfur hydride, but still is rather high.

The idea combining high pressure and doping is elegant and looks promising. One has to wait whether the future experiments will confirm this interesting prediction.

### D. Organic hydrides

The presence of hydrogen and corresponding high vibrational frequency is very beneficial for the formation of the superconducting state. As was noted in the previous section, it can be manifested even at ambient pressure. In connection with this, the recent discovery of superconductivity in the organic compound, consisting of C and H elements, so-called poly(p-phenylene) and doped with potassium (Wang et al.,2017) looks very promising.  The value of $T_c$ is rather high ($T_c \approx 120K$); this value is the highest among organic superconductors. Organics is a relatively young family of superconducting materials. The first organic superconductor was discovered in 1980 by  D. Jerome:

the complex material $(TMTSF)_2PF_6$ displayed the property at $T_c \approx 0.9K$ under the pressure of around 9 kbar. The recent discovery was made by Wang et al. in the high pressure laboratory, but the effect was observed at ambient pressure. The mechanism of superconductivity in this new material and an impact of the hydrogen bonding and high frequency modes should be studied in full details, As a whole, this new class of organic hydrides looks very important and deserves a further study.

### VIII. Main challenges.

The recent discovery of the record-breaking high $T_c$ compound, sulfur hydride, signifies the arrival of a novel family of high temperature superconductors: the hydrides. Even higher values of $T_c$ can be expected. That is, it finally becomes perfectly realistic to envision the detection of superconductivity at room temperatures. The search for novel hydrides with still higher values of $T_c$ (including $CaH_6$, $YH_6$) is a very important direction of future research.

More detailed X-ray diffraction of sulfur hydrides will establish the position of the sulfur atoms with high accuracy and clarify the nature of the phase transition between the "low $T_c$" and "high $T_c$" phases.

Development of tunneling spectroscopy of the high $T_c$ hydrides is another important forthcoming task. Because of the large values of $T_c$ and the energy gap, the tunneling I(V) characteristic needs to be measured for a wider energy interval as compared to that for conventional superconductors. But similar measurements have been performed for the high $T_c$ cuprates by Aminov et al. (1994) and Ponomarev et al. (1999) with the use of the break junction technique, and by Lee et al. (2006) by scanning tunneling spectroscopy, see review by Kresin and Wolf (2009) . Therefore tunneling spectroscopy should be successfully applicable for sulfur hydrides as well. As a result, it will be possible to reconstruct the $\alpha^2(\Omega)F(\Omega)$ function as well as to determine the Coulomb pseudopotential $\mu^*$. Tunneling spectroscopy also can be employed for measuring other important parameters of the system, including the energy gap, and for observing the multi-gap structure and its evolution with pressure (see Ch. VI).

And, of course, following detailed studies of the structure of the high $T_c$ phase under pressure, there remains the most intriguing question: is it possible to create analogous structures stable at ambient pressure?

### IX. Concluding remarks.

The discovery of pressure-induced superconductivity in the hydride family opens new prospects in research on high temperature superconductivity. In this Colloquium we have focused mainly on the theoretical aspects of this new development.

From the fundamental point of view, it is remarkable that the high $T_c$ superconducting state manifests itself under such extreme experimental conditions. Sulfur hydrides offer a remarkable combination of strong electron-phonon coupling and high optical-phonon frequencies. As long as the Migdal adiabaticity criterion is not violated, observations of superconductivity at even higher temperatures now can be anticipated.

There are, however, novel features, which necessitate significant deviations from the conventional Migdal-Eliashberg approach. First of all, the hydride phonon spectra are quite broad (up to 200 $meV$) and contain both optical and acoustic modes. We propose that the electron-phonon interaction can be treated with the use of the general equation (4.1) by employing two coupling constants $\lambda_{opt}$ and $\lambda_{ac}$ together with two corresponding average frequencies. This leads to an analytical expression for $T_c$ applicable to a number of cases, and permits analysis of other relevant factors for different phases.

The experimentally observed isotope effect (deuterium-hydrogen substitution) turns out to be not universal in the sense that the isotope coefficient depends on the pressure and has distinct values for different phases. We show that this reflects the relative contributions of optical and acoustic modes.

We point out that the sharp increase in $T_c$ (from $\approx 120K$ up to $\approx 200K$ over a narrow pressure interval near $\approx 150 GPa$) is a signature of a first-order structural transition into the high $T_c$ phase. This picture also explains the curiously slow decrease in $T_c$ at $P>P_{cr.}$

It has been predicted in a number of theoretical papers that the transition into the high $T_c$ phase is accompanied by the appearance of additional small Fermi pockets. As a consequence, a two-gap structure appears . It should be observable by tunneling spectroscopy, which also will be useful for determination of the characteristic function $\alpha^2(\Omega)F(\Omega)$.

It may be expected that there should exist other hydrides capable of displaying a high $T_c$ superconducting state under pressure, possibly with even higher values of $T_c$ all the way up to room temperature. But the most challenging question relates to the possibility of creating superconducting structures stable at ambient pressure. In this regard, we are encouraged by the recent observation by Syed et al.,2016, of the superconducting state of Pd-H at temperatures ≈54 K and ≈60 K for Pd-D.

**Acknowledgements**

The authors thank M. Eremets, A. Drozdov, M. Calandra, M.Einaga, and Y.Ma for interesting and stimulating discussions. The work of LPG is supported by the National High Magnetic Field Laboratory through NSF Grant No. DMR-1157490, the State of Florida and the U.S. Department of Energy. The work of VZK is supported by the Lawrence Berkeley National Laboratory, University of California at Berkeley, and the U.S. Department of Energy.